\documentstyle[aps,prl,preprint]{revtex}
\tightenlines
\begin{document}
\draft
\title{Diffusion of Point Defects in Two-Dimensional Colloidal Crystals}
\author{Alexandros Pertsinidis\cite{coresp} and X. S. Ling}
\address{Department of Physics, Brown University, Providence, Rhode Island 02912}

\date{\today}
\maketitle
\begin{abstract}
We report the first study of the dynamics of point defects, mono and di-vacancies, in a confined 2-D colloidal crystal in real space and time using digital video microscopy. The defects are introduced by manipulating individual particles with optical tweezers. The diffusion rates are measured to be $D_{mono}/a^{2}\cong3.27\pm0.03$Hz for mono-vacancies and $D_{di}/a^{2}\cong3.71\pm0.03$Hz for di-vacancies. The elementary diffusion processes are identified and it is found that the diffusion of di-vacancies is enhanced by a \textit{dislocation dissociation-recombination} mechanism. Furthermore, the defects do not follow a simple random walk but their hopping exhibits memory effects, due to the reduced symmetry (compared to the triangular lattice) of their stable configurations, and the slow relaxation rates of the lattice modes.
\end{abstract}

Uniform colloidal spheres suspended in a solvent, under appropriate conditions can self-assemble into ordered crystalline structures \cite{colloid_crystals}. Using these crystals as a model system, a great variety of problems in materials science, physical chemistry and condensed matter physics have been investigated during the past two decades. However, the properties of simple structural defects, such as vacancies and interstitials, dislocations, etc., ubiquitous in real solids, have received little, if any, attention. Considering the unique possibility of real space and real time imaging, colloidal systems provide a great potential for research in this direction.

Recently, it has been demonstrated\cite{Part_A} that point defects can be created in two-dimensional colloidal crystals\cite{2D_colloid} by manipulating individual particles with optical tweezers. Direct imaging of these defects verified that their stable configurations have lower symmetry than the underlying triangular lattice, as predicted by numerical simulations for a number of two-dimensional systems\cite{w_xtal,defect_configs}. The same study revealed that point defects can dissociate into pairs of well-separated dislocations, a topological excitation permitted only in two but not three dimensions. In this Letter we report the first study of the dynamics of mono and di-vacancies in two-dimensional colloidal crystals. We discover the first evidence that topological excitations into dislocation pairs enhance the diffusion of di-vacancies. Moreover, the hopping of the defects does not follow a pure random walk, but exhibits surprising memory effects.

The colloidal crystals are prepared by 0.360 $\mu$m diameter negatively charged polystyrene-sulfate micro spheres (Duke Scientific, \#5036), suspended in highly deionized water. The particles crystallize due to strong electrostatic interaction. A single layer of particles is confined between two fused silica substrates separated by $\approx$2 $\mu$m, resulting in a single layer colloidal crystal with a lattice constant $a\cong1.1 \mu$m. The experimental details can be found in\cite{Part_A}. Point defects are created by trapping a particle with optical tweezers and dragging it away from its lattice site.
The dynamics of such single, isolated point defects are recorded on videotape, using a CCD video-camera. Each individual frame (1/60th of a second) was acquired on a PC and processed using a particle-tracking algorithm\cite{prtcl_track}. A triangulation algorithm is used to identify the mis-coordinated particles and characterize the configuration of the defect.

We tracked a number of mono and di-vacancies for long enough times (up to 40 seconds) to obtain reliable measurements of their diffusion constants. Typical trajectories are shown in Fig.\ref{mono_traj}. The position of the defect $\vec{x}(t)$ is defined as the average of the positions of the 5-fold coordinated particles around the \textit{core} of the defect. $\vec{x}(t)$ is determined with a resolution of $\approx 0.1a$. Thus one can identify the defect in a particular lattice site or in-between two neighboring sites. To estimate the diffusion constants, we take the average of the squared displacements $<\Delta \vec{x}^{2}(\delta t)>$ as a function of the time separation $\delta t$, running through all the available trajectories. A linear relationship $<\Delta \vec{x}^{2}>$ vs. $\delta t$ is obtained for the range 1-5 seconds (Fig.\ref{mean_x2}\textbf{A}). The data in this range are weighted by their errors and fitted to a zero y-intercept linear relation, whose slope gives the respective diffusion constants. The results for mono and di-vacancies are: $D_{mono}/a^{2}\cong3.27\pm0.03$Hz, $D_{di}/a^{2}\cong3.71\pm0.03$Hz\cite{E_barrier}.

The diffusion constants measured above give the combined contribution of all the possible diffusion pathways. In this Letter, we reveal the role of topological excitations on the diffusion of point defects. As was shown in \cite{Part_A}, the possible stable configurations of mono and di-vacancies are not just the "well-identified" point defect configurations but there are also configurations where the point defects appear as separated dislocation pairs. 
The defects are observed to dissociate into two dislocations that glide from site to site and finally recombine at a new position that can be several lattice constants away from the original one. The individual dislocations move faster than the hopping of mono or di-vacancies and this topological excitation has a \textit{lifetime} $\tau_{o}$ of a fraction of a second. Such a process is a distinct diffusion pathway, giving a contribution to the diffusion constant $D_{pair}\approx\tau_{o}^{-1}l^{2}$, where $l^{2}$ is the mean square separation of the dislocation pair.

Mono-vacancies spend considerably less time as a dislocation pair than di-vacancies and the aforementioned diffusion mechanism is not important in their case. One can clearly recognize the different behavior exhibited by mono and di-vacancies, as illustrated in Fig.\ref{separation}, where the separation of the dislocations comprising the defect is shown as a function of time. The mono-vacancy is observed primarily to fluctuate between configurations in which the dislocations are bound, with very rare excursions into configurations where the dislocations dissociate. In contrast, the di-vacancy frequently spends long intervals (a few seconds) as a dislocation pair, with the two dislocations dissociating and recombining several times during this period\cite{ddr_mech}.

The combined diffusion constant for di-vacancies is $D_{di}=pD_{pair}+(1-p)D^{o}_{di}$, where $p\approx0.25$ is the fraction of time excited into a dislocation pair, and $D_{pair}$,$D^{o}_{di}$ are the contributions from the dissociation-recombination mechanism and from the "bare" di-vacancy hopping. In order to estimate the diffusion constants for the two different mechanisms, we link into two separate trajectories the intervals spent as a dislocation pair and as a "bare" di-vacancy. To avoid systematically picking up hoping events, the shortest interval was 1 second (larger than the time between jumps). Carrying out the same procedure as for the original trajectories, the two diffusion constants are found to be $D_{pair}/a^{2}=5.90\pm0.09$Hz and $D^{o}_{di}/a^{2}=2.97\pm0.03$Hz (Fig.\ref{mean_x2}\textbf{B}). These numbers combine to give the measured value $D_{di}/a^{2}=3.70$Hz.

We now turn our attention to the details of the underlying stochastic hopping processes. Naively, one would expect the trajectories of the defects to be pure random walks on a triangular lattice, i.e. when hopping, the system chooses with equal probability between the six neighbor sites and consecutive events are uncorrelated. However, a careful analysis reveals a peculiar feature: their motion looks nothing like a 2D random walk, since typically it is restrained along particular directions\cite{Auxiliary}. The existence of \textit{easy} directions for diffusion is related to the low symmetry of the stable configurations of the relaxed lattice around the defect\cite{Part_A,defect_configs}. Therefore, the probability for hopping towards the six nearest neighbors is not the same for all of them, but is higher along particular lines of symmetry, Fig.\ref{easy_directions}. For di-vacancies excited into a pair of separated dislocations, the easy direction coincides with the gliding direction of the two dislocations. The easy direction for diffusion remains unchanged for up to a few seconds.

The existence of easy directions for hopping makes the trajectories of the defects quasi-one-dimensional, at least for a few consecutive steps. This effect is demonstrated by measuring the probability $P_{o}(n)$ for returning to the original position after the defect has executed $n$ steps. It is found that the measured $P_{o}(n)$ for small $n$ resembles the one for random walk on a 1-D chain, rather that for random walk on a 2-D triangular lattice. This is illustrated in Fig.\ref{Po(n)}: $P_{o}(n)$ clearly deviates from the expected 2-D behavior, following the characteristic oscillations of the 1-D random walk of period=2 steps.

A second surprising peculiarity involves memory during site-to-site hopping. The defects appear to spend some time around a lattice site, attempting to hop towards neighboring sites by changing into a split intermediate configuration between the two lattice points. Depending on the symmetry of the relaxed lattice in the original configuration (2 or 3-fold), the defect is seen to attempt hopping towards either two or three of the six neighbors. Oddly, the defect typically hops back to the original site, finally getting to the next site \textit{only after several attempts}. The back-hopping to forward-hopping ratio is found to be $\approx3:1$.

This \textit{memory effect} originates from the slow relaxation rates of distotrions in the overdamped colloidal crystal: the core region of the defect can rearrange rapidly ($\sim10Hz$), however the rest of the lattice around the defect cannot follow at the same rate (e.g. distortions with wavelength $10a$ relax at $\approx1Hz$). As a vacancy changes from a configuration at one site to a split intermediate configuration between two sites, the rest of the lattice does not immediately relax, and seems to "push" the vacancy back to its initial site\cite{memory}.

In summary, the dynamics of point defects in colloidal crystals are studied in real space and time. Di-vacancies in particular are observed to dissociate into two dislocations that can separate several lattice spacings before recombination. This distinct mechanism enhances the diffusion rate of di-vacancies compared to their "bare" one. The reduced symmetry of the defects' stable configurations results in \textit{easy} directions for diffusion. We have also found that the site-to-site hopping seems to exhibit memory effects accounted for by the slow relaxation of the short wavelength modes of the crystal.
We acknowledge helpful discussions with Prof. S.C. Ying and Prof. D.A. Weitz.
This work was supported by NSF Grant No. DMR-9804083, the Petroleum Research Fund, the Research Corporation, and the A.P. Sloan Foundation.


%
%

%
%

\begin{figure}
\caption{Trajectories of defects: (Top) mono, (Bottom) di-vacancy. Neighboring \textit{clumps} of dots are separated by roughly half a lattice spacing and correspond to the defect being in a configuration near a lattice point or in a split configuration between two lattice points. A snapshot of the lattice is super-imposed a few seconds after the defect diffused away from the region shown. The positions of the defect are displaced with respect to the lattice seconds later due to slow, long wavelength, thermally generated distortions.}
\label{mono_traj}
\end{figure}

\begin{figure}
\caption{Top: Mean square displacements of mono and di-vacancies. The average of $\Delta x^2$ as function of $\Delta t$ was measured from about 2-minute long mono and di-vacancy trajectories. The straight lines are linear fits for $\Delta t$ 1-5 seconds. Bottom: $\Delta x^2$ separately for "bare" di-vacancies and di-vacancies excited into dislocation pairs. Fits are for $\Delta t$ 1-3 seconds.}
\label{mean_x2}
\end{figure}

\begin{figure}
\caption{Separation $r(t)$ of the dislocations comprising a mono and a di-vacancy (top and bottom respectively). $r(t)$ is measured as the standard deviation of positions of the 5-fold coordinated particles in the defect core. The positions of the minima in the dislocation pair potential\protect\cite{Part_A} are indicated by the dotted horizontal lines. The first two correspond to the split and the crushed configurations, while above the third one (emphasized with a dashed line) the dislocation pair is dissociated. Note that in the mono-vacancy case a large fraction of the points shown (between $SV$ and $V_{2}$) corresponds to the $V_{3}$ configuration, which is a bound dislocation triplet. In the case of di-vacancy, the arrows indicate intervals in which the defect appears as a dislocation pair. The two dislocations repeatedly dissociate and recombine during this period.}
\label{separation}
\end{figure}

\begin{figure}
\caption{Configurations of mono-vacancy, from\protect\cite{Part_A}. Top: three-fold symmetric $V_{3}$, bottom: crushed $V_{2a}$. The 7-fold coordinated particles are more probable to hop towards the vacant site, resulting in particular \textit{easy} directions for diffusion (indicated with arrows).}
\label{easy_directions}
\end{figure}

\begin{figure}
\caption{$P_{o}(n)$ measured from 40 seconds of data or roughly 60 hopping events. Clearly, the observed behavior for both type of defects resembles more the one for 1-D random walk, rather the one for 2-D. $P_{o}(n)$ exhibits oscillations of period 2 steps and the average trend lies between the 1-D and 2-D curves.}
\label{Po(n)}
\end{figure}


\begin{references}
\bibitem{ddr_mech} A plausible explanation for this unexpected behavior is the following: (i) the energy barrier for dissociation of a mono-vacancy is about $2k_{B}T$ larger than the one of a di-vacancy\cite{Part_A}. (ii) The dissociation of a point defect into a dislocation pair originates from a configuration with two-fold symmetry. Di-vacancies naturally prefer such configurations. On the other hand, mono-vacancies spend considerable amount of time in a three-fold configuration. As a result, the fraction of time a mono-vacancy spends excited into a dislocation pair is $p\approx 0.01$.
\bibitem{memory} This memory could influence the dynamics of the defects at longer times as well. Such an effect would be manifested in e.g. the behavior of the return probability $P_{o}(n)$ (Fig.\ref{Po(n)}), however in a manner unclear to us at this point.
\end{references}
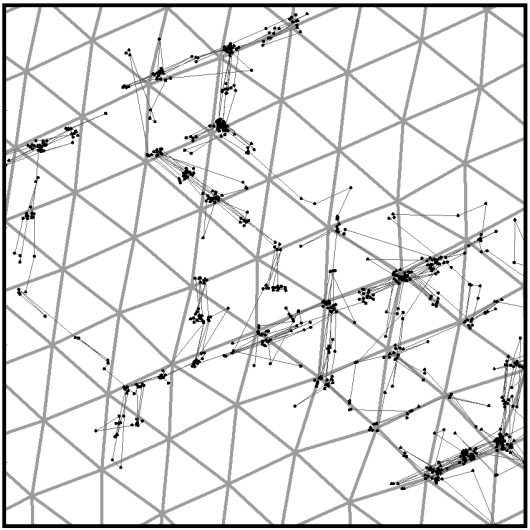
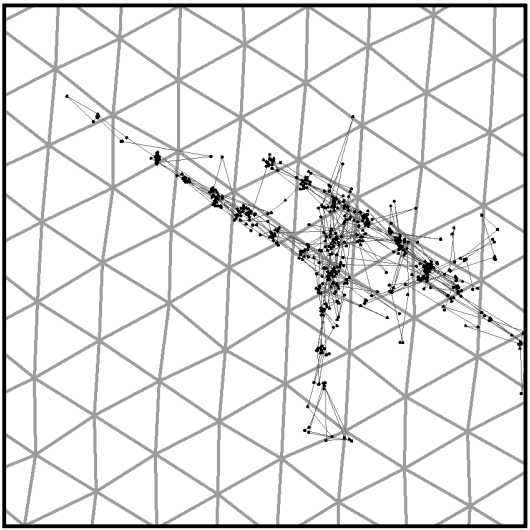
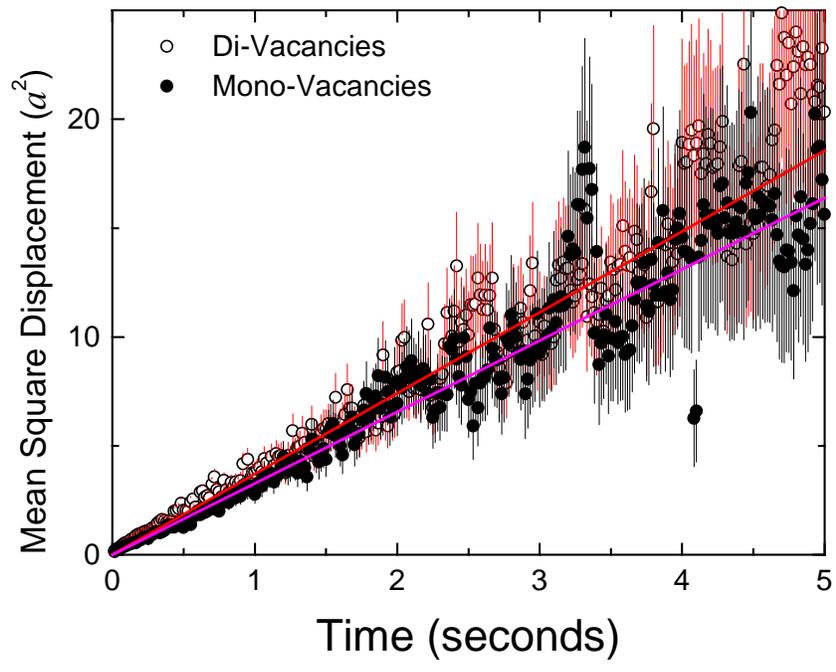
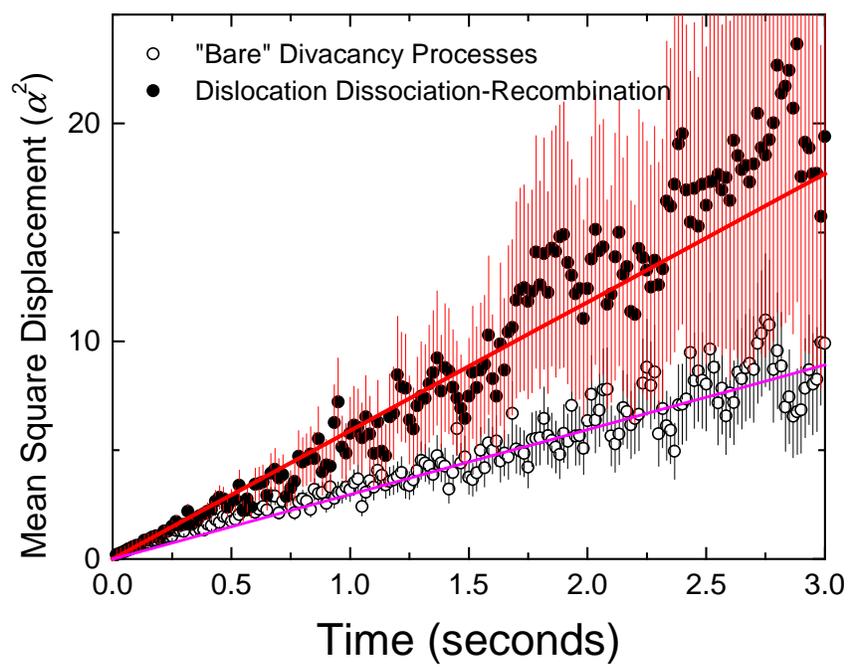
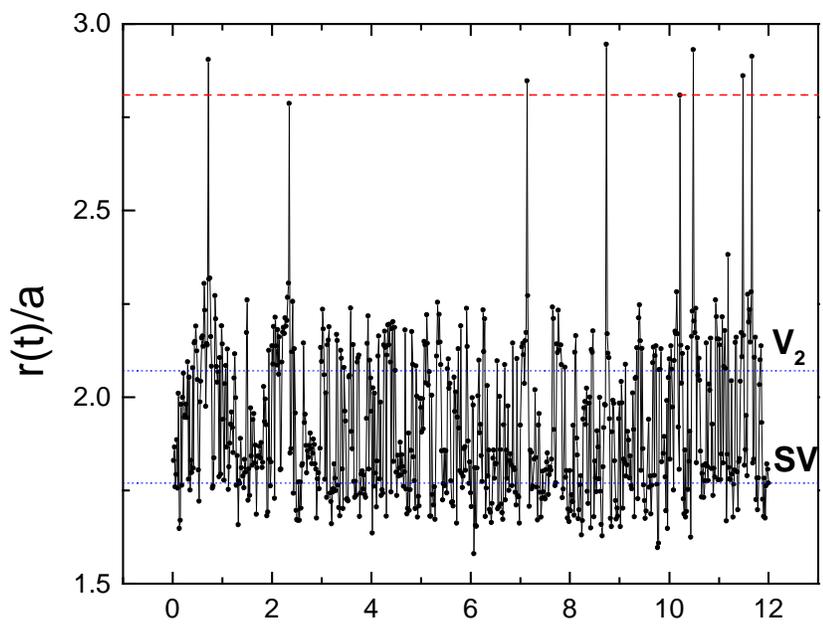
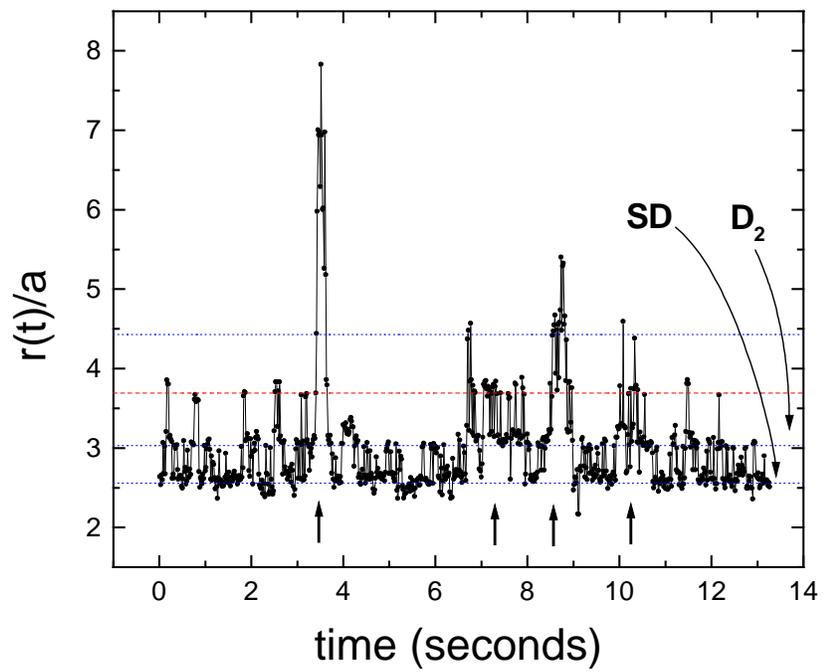
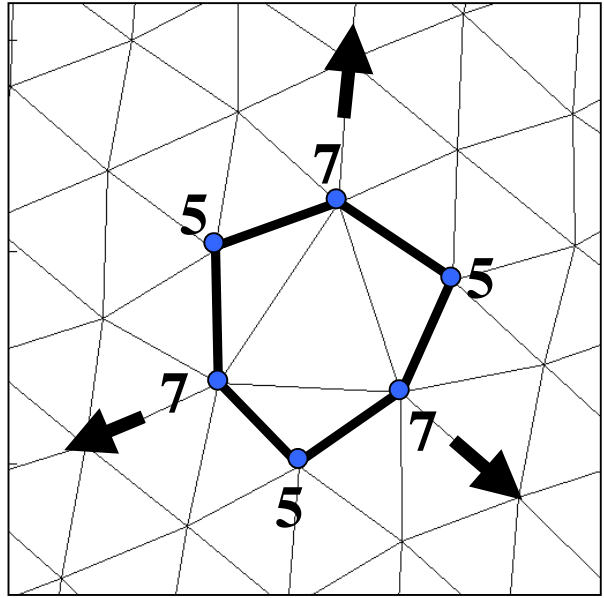
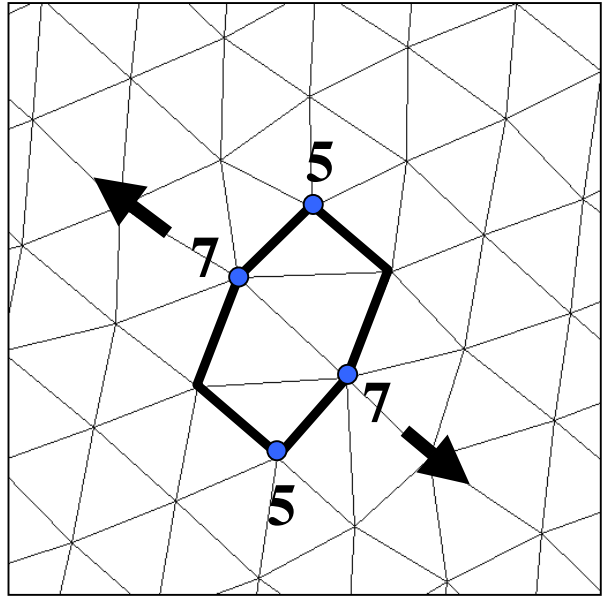
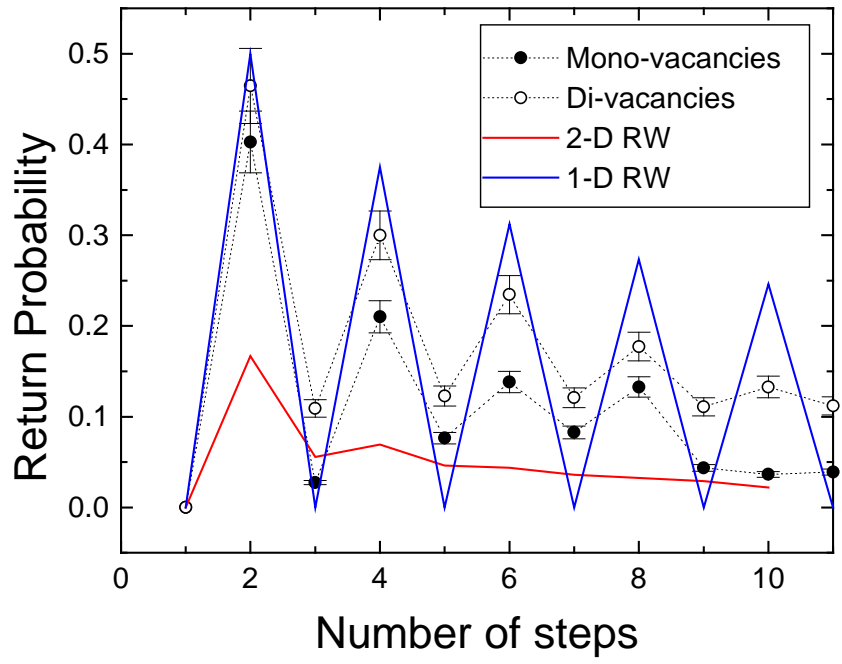
\end{document}